# An Ontology Based Modeling Framework for Design of Educational Technologies

**Sridhar Chimalakonda · Kesav V. Nori**

**Abstract** Despite rapid progress, most of the educational technologies today lack a strong instructional design knowledge basis leading to questionable quality of instruction. In addition, a major challenge is to customize these educational technologies for a wide range of instructional designs. Ontologies are one of the pertinent mechanisms to represent instructional design in the literature. However, existing approaches do not support modeling of flexible instructional designs. To address this problem, in this paper, we propose an ontology based framework for systematic modeling of different aspects of instructional design knowledge based on domain patterns. As part of the framework, we present ontologies for modeling *goals*, *instructional processes* and *instructional materials*. We demonstrate the ontology framework by presenting instances of the ontology for the large scale case study of adult literacy in India (287 million learners spread across 22 Indian Languages), which requires creation of 1000 similar but varied *e*Learning Systems based on flexible instructional designs. The implemented framework is available at http://rice.iiit.ac.in and is transferred to National Literacy Mission of Government of India. This framework could be used for modeling instructional design knowledge of systems for skills, school education and beyond.

**Keywords** scale; variety; educational technologies; ontologies; families; instructional design; goals; instructional process; content; *e*Learning Systems

S. Chimalakonda
Department of Computer Science & Engineering
Indian Institute of Technology, Tirupati
E-mail: ch@iittp.ac.in

K.V. Nori
Software Engineering Research Center
International Institute of Information Technology Hyderabad, India
E-mail: ch@iittp.ac.in

## 1 Motivation

John McCarthy has envisioned that an intelligent way of building systems should focus on the *knowledge* that is required to represent system's inputs and methods through which possible conclusions can be automatically derived from that knowledge (McCarthy, 1963). Newell has proposed the need to have a *knowledge level* focusing on specifying the world independent of *symbol level* that focuses on implementing the behaviour of the system (Newell, 1982).

> "The Knowledge Principle: A system exhibits intelligent understanding and action at a high level of competence primarily because of the *specific knowledge* that it can bring to bear: the *concepts*, *facts*, *representations*, *methods*, *models*, *metaphors*, and *heuristics* about its *domain* of endeavor." -Lenat and Feigenbaum (Lenat & Feigenbaum, 1991)

Feigenbaum coined the term *knowledge engineering* and proposed knowledge base should be a fundamental basis that stores expertise of human experts in solving real-world problems (Lenat & Feigenbaum, 1991). There are two primary directions of research related to knowledge engineering one in the field of Artificial Intelligence primarily for automatic reasoning and expert systems and in computer science to represent different aspects of the system. In this paper, we are interested in modeling *knowledge* in the domain of education to facilitate design of educational technologies[1] and *e*Learning Systems[2]. We are specifically interested in the following questions:

– *What is the knowledge that is required to facilitate the design of educational technologies for scale and variety? where scale is the number of systems to be developed and variety represents the different kinds of systems to be developed in education domain.*
– *How do you concretely represent this knowledge?*

Towards answering these questions, the core contribution of this paper[3] is:

– *An ontology based framework to model instructional design knowledge to facilitate scale & variety during design of educational technologies*
– *The proposed framework is evaluated by demonstrating ontologies for the large scale case study of adult literacy throughout the paper*

The rest of the paper is as follows: A brief background of adult literacy case study is presented in Section 2. Existing work on ontologies for instructional design and for adult literacy are discussed in Section 3 as part of related work. A broad spectrum of ontologies and their development process is in Section 4. In Section 5, we present our ontology based framework for modeling instructional design. Within this section, we detail ontologies for modeling goals, instructional process and instructional material in sub sections 5.1, 5.2 and 5.3. We present a concrete instance of domain ontology for adult literacy as a instantiation of the ontology in Section 6.1. Finally, we end the paper with conclusions and future work in Section 7.

[1] We consider *"educational technologies as a set of processes, techniques, methods and tools that facilitate systematic development of eLearning Systems based on well-established instructional designs."*

[2] We consider *"eLearning Systems as a sub-class of educational technologies that are designed for improving learning and teaching in a particular context."*

[3] This paper is an extensively revised version of our earlier three page short paper that was published during its formative stages (Chimalakonda & Nori, 2013) and from the doctoral thesis of first author (Chimalakonda, 2017).

## 2 Background - A Case Study

### 2.1 The Scale & Variety Challenge

There are 287 million adult illiterates in India spread across 22 Indian Languages who can speak the language, but cannot read or write (*Education for All Global Monitoring Report 2013/4: Teaching and learning: Achieving quality for all*, 2014). The National Literacy Mission of Government of India has come up with a uniform methodology called Improved Pace and Content of Learning (IPCL) to teach adult illiterates across India (*Handbook for Developing IPCL Material*, 2003). Based on this methodology, nearly 1000 primers[4] are created and further customized for 22 Indian Languages. This presents the need to develop nearly 1000 *e*Learning Systems or iPrimers[5] and further customize them for 22 Indian languages. We discussed this challenge of *scale* and *variety* in the context of adult literacy in India at length as part of first author's doctoral thesis (Chimalakonda, 2017). Essentially, the goal of designing and customizing educational technologies presents the need to model *knowledge* pertaining to different aspects of instructional design so as to facilitate *scale* and *variety*. We rely on (i) IPCL, a pedagogy and a process for teaching 3Rs (Reading, wRiting, aRithmetic) to adult illiterates and which provides guidelines to prepare instructional materials across multiple languages and varied contexts (*Handbook for Developing IPCL Material*, 2003) (ii) field-tested *e*Learning systems based on this primers (iii) domain specific patterns of instructional design (Chimalakonda & Nori, 2014) (Chimalakonda, 2017). We are concerned with the following questions:

- *What is the knowledge that is required to automate the development of eLearning systems for scale and variety in the context of adult literacy in India?*
- *How can we represent this knowledge in order to customize these eLearning Systems for flexible instructional designs?*

### 2.2 Concrete Knowledge in Adult Literacy Case Study

We briefly present a concrete example of knowledge from adult literacy case study as a first step for setting the context for rest of the paper. The IPCL

[4] Primers are essentially printed textbooks based on customized instructional designs.

[5] We consider *eLearning Systems as simple multimedia systems that use audio and visual aspects to teach reading, writing and basic arithmetic corresponding to physical instructional material (Chimalakonda, 2010)*

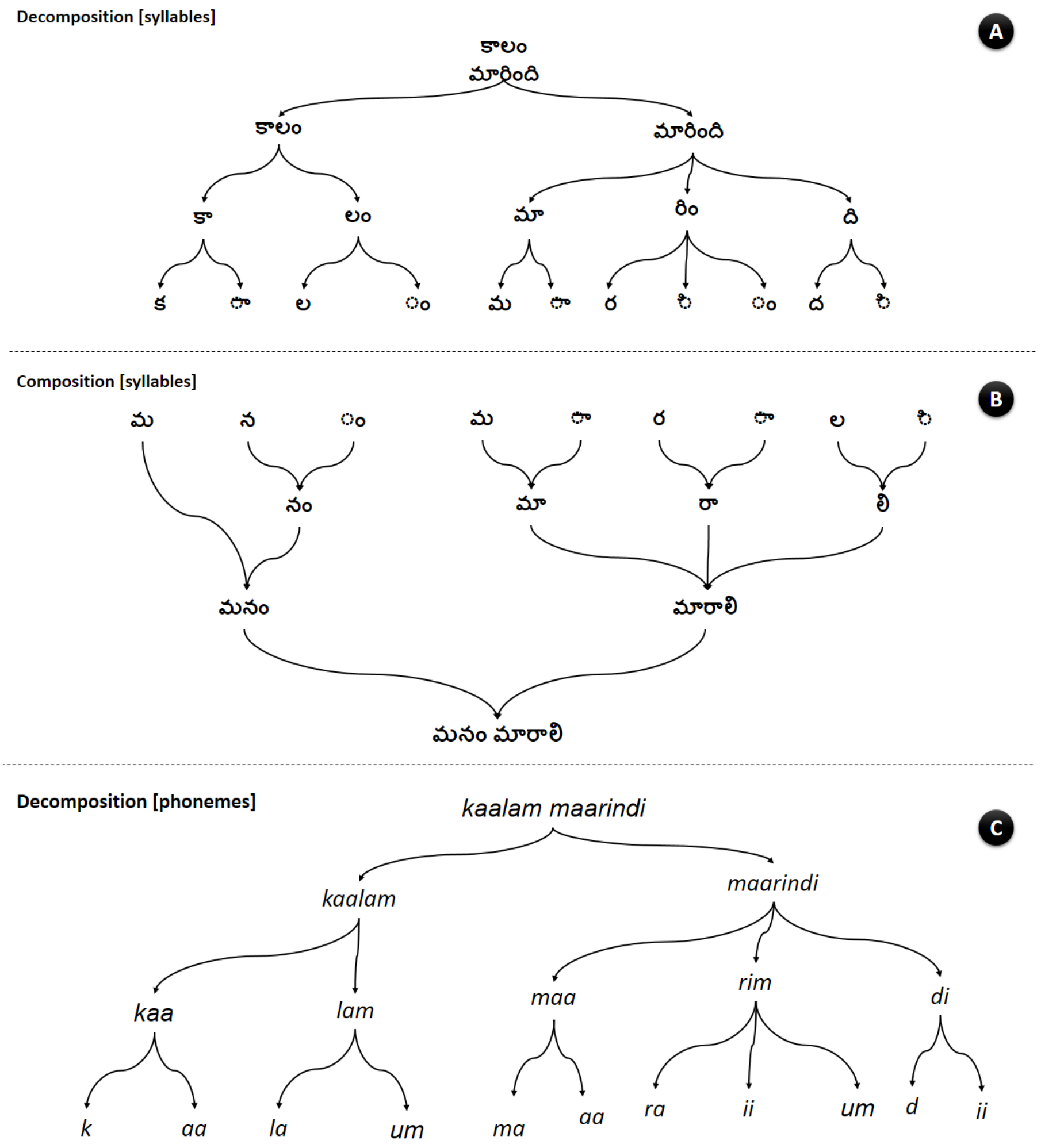

**Fig. 1** Top-down decomposition of sentences to syllables [A], Bottom-up composition of words and sentences from syllables [B], Top-down decomposition of sentences to phonemes [C]

approach suggests the use of *eclectic method* for teaching reading skills, comprehension, problem solving and facilitates learning through interpretation of contents in the context of life (*Handbook for Developing IPCL Material*, 2003)(*Confintea VI: sixth international conference on adult education: final report*, 2010). We consider "context of life" as learners' prior knowledge. Ecletic method primarily differs from traditional methods as it does not start with alphabets but instead uses familiar and known words to learners, decomposes them to syllables and phonemes as shown in Figure 1[A][C].

These syllables and phonemes are further synthesized to form words and in the end, the alphabet is learnt as in Figure 1[B]. The patterns of decomposition (top-down) focuses on cognitive abilities of learners whereas the patterns of composition (bottom-up) can facilitate reasoning of the subject knowledge. We have depicted examples of this process in Figure 2 for the *Telugu* language based on the primer (instructional material). In this figure, a sentence in *Telugu* language కాలం మారింది is first decomposed into two words namely కాలం, మారింది. Each of these words are further decomposed till the syllables are obtained. Similarly, the same sentence is also decomposed into phonemes representing the sounds of the sentence, words and syllables respectively.

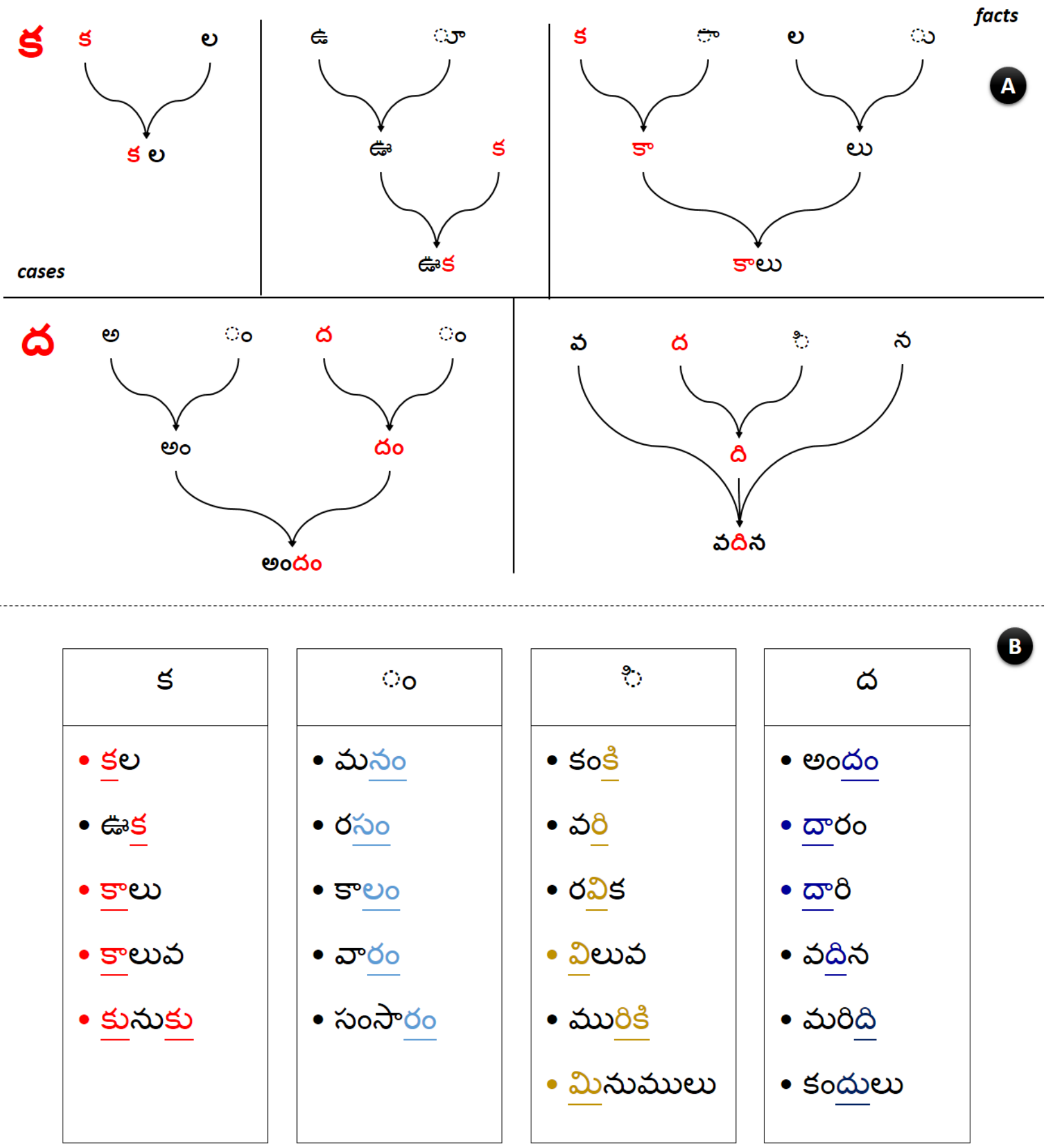

**Fig. 2** Composition of cases from facts [A], Possible set of cases for a lesson in *Telugu* language primer [B]

On the other hand, Figure 2[A] shows the composition process which uses the syllables and phonemes to form words and sentences. Specifically, words

కల, ఊక and కాలు are formed by composing the syllable క and other relevant syllables and words అందం, వదిన are formed from their respective syllables with the learnt syllable ద in red color. This hierarchy of decompositions and compositions forms the basis for learning new syllables and phonemes in a language as shown in Figure 2[B]. Several compositions are possible V-Vowel+Vowel Modifier, C-Consonant+Vowel Modifier, C+V-Consonant+Vowel, Vowel Modifier, C+C-Consonant+Consonant, C+C+V-Consonant+Consonant+Vowel and there will be several constraints on these compositions as well. For example, only one modifier might be allowed after a consonant. Discussing that in detail is not within the purview of this paper. But this approach of learning alphabet from known words and sentences and later synthesizing new words from syllables and phonemes was empirically established by IPCL as a successful approach for adult literacy in India (*Handbook for Developing IPCL Material*, 2003). Most importantly, this approach works for the scale and variety of 22 Indian languages and varied instructional designs. This specific knowledge has been abstracted into instructional design knowledge and patterns (Chimalakonda, 2017). Generalizing from this specific knowledge, the requirements for knowledge representation in the context of this paper are as follows. The representation should be:

- in synergy with instructional design
- machine-processable
- facilitate reuse and semi-automatic design of *e*Learning Systems
- able to support sharing of knowledge between different applications and tools

In the next section, we discuss related work for modeling instructional design knowledge in the context of this paper.

## 3 Related Work

Several researchers have figured out multiple ways of representing knowledge like concept maps, topic maps, ontologies, first order logic and so on (Sowa, 1999)(Baral & De Giacomo, 2015).

### 3.1 Ontologies for Instructional Design

Ontologies have gained immense importance in the last few decades as one of the widely used methods to represent and share *knowledge* in several domains such as software engineering (Happel & Seedorf, 2006), enterprise modeling (Pinto, de Rezende Rohlfs, & Parreiras, 2014), requirements engineering (Dermeval et al., 2015). These ontologies are of different kinds ranging from informal light weight ontologies to formal ontologies depending on the degree of formalism and the power of expressivity (Giunchiglia & Zaihrayeu, 2009). Happel et al. have discussed the advantages of ontologies over conceptual models and meta-models (Happel, Maalej, & Seedorf, 2010) as follows:

- Enable new and efficient way to information reusability.
- Enable to extend easily.

- Provide consensus on the understanding of domain knowledge.
- Support better understanding of domain knowledge.
- Define problem and solution domain knowledge separately.
- Assist in analyzing the structure of domain knowledge.
- Facilitate a machine to use the knowledge in an application.
- Share common semantics among people and applications.

Fensel attributes the popularity of ontologies is due to the promise of providing *"a shared and common understanding of a domain that can be communicated between people and application systems"* (Fensel, 2001), which can be construed as *representation*, *communication* and *automation* needs for scale and variety in the design of educational technologies.

In the domain of education, ontologies are extensively used (Mizoguchi & Bourdeau, 2000)(Mizoguchi & Bourdeau, 2015) in a wide range of applications ranging from explicit representation of domain knowledge to automatic generation of personalized content (Sampson, Lytras, Wagner, & Diaz, 2004). Mizoguchi and Bourdeau have identified four key requirements of instructional authoring systems (i) adaptivity (ii) explicit conceptualization (iii) standardization to facilitate reuse (iv) theory-awareness, and proposed knowledge and ontological engineering as a potential solution to cater to these requirements (Mizoguchi & Bourdeau, 2000).

One particular use of ontologies that is of interest to this paper is to model instructional design theories and learning designs (Psyché, Bourdeau, Nkambou, & Mizoguchi, 2005) but during design of educational technologies. A 10-year research effort has resulted in creating a comprehensive ontology covering instructional design knowledge for various instructional theories and adhering to learning design standards (Mizoguchi, Hayashi, & Bourdeau, 2007). SMARTIES is a scenario-based instructional authoring tool based on this ontology and advocates the design of educational technologies based on educational theories modeled as ontologies to facilitate quality of instruction. However, the inherent complexity of the ontology and SMARTIES tool made it tough for its practical usage (Kasai, Nagano, & Mizoguchi, 2011).

While focusing on quality of instruction is one aspect, using ontologies in education to facilitate reuse is another critical research direction that received significant attention in the literature (Mizoguchi & Bourdeau, 2000). Devedzic explored the notion of ontologies for intelligent tutoring systems (ITS) based on inspiration from software patterns in 1999 (Devedzic, 1999). Ontologies to formalize learning object content models have been proposed in (Verbert, Klerkx, Meire, Najjar, & Duval, 2004). To facilitate flexible content reuse, the Abstract Learning Object Content Model (ALOCoM) ontology and a set of supporting tools were proposed in (Verbert et al., 2004). Amorim et al. have proposed a learning design ontology based on IMS LD specification through a set of 20 design and run time axioms (Amorim, Lama, Sánchez, Riera, & Vila, 2006). The basic premise of this ontology was to explicitly and precisely address the drawbacks of IMS LD specification (Amorim et al., 2006). But isolated research on learning objects and learning designs have made reuse dif-ficult motivating the need for a bridge ontology focusing on context (Knight, Gasevic, & Richards, 2006). A formal ontology was presented for representing instructional design methods and provides a rule catalogue to verify the conformance of ontologies for a particular instructional design theory (Vidal-Castro, Sicilia, & Prieto, 2012). An ontology and a software framework focusing on competencies was discussed in (Paquette, 2007). However, the creation of these ontologies is not based on domain-specific patterns (Chimalakonda, 2017), which is the case in this paper. Furthermore, the existing ontologies are not aimed at *scale* and *variety*, which is an essential goal of this paper.

There has been a significant growth of research in ontologies in the past decade resulting in a variety of approaches for ontological engineering (N. F. Noy, 2004)(Gomez-Perez, Fernández-López, & Corcho, 2006)(Dicheva, 2008) (Pinto et al., 2014)(Mizoguchi & Bourdeau, 2015)(Dermeval et al., 2015). In particular, ontologies have been used for representing different aspects of a specific domain using a wide range of different mechanisms (Uschold & Gruninger, 2004). At the core of ontologies is the identification of concepts generally represented as a hierarchy of classes and sub-classes and different relationships between them. Each of the these classes typically have associated properties and can also have a set of constraints. Instances of these ontologies are called as individuals and represent a specific instance of a particular domain represented by the ontology. For example, an ontology for instructional design at a higher level can be defined using concepts like *goals*, *process*, *content* and so on and each of these sub-classes can be further defined. An instance of this ontology can be a specific instructional design for teaching a specific course. We discuss these ontologies in detail in the rest of the paper.

Diversified needs emerging from different domains gave raise to a spectrum of ontology kinds (Uschold & Gruninger, 2004)(Giunchiglia & Zaihrayeu, 2009) as shown in Figure 3. These kinds of ontologies vary based on the degree of specification detail, formalism and expressiveness power as we move from one end to the other end of the spectrum. A detailed description of this spectrum is given in (Uschold & Gruninger, 2004)(Wong, Liu, & Bennamoun, 2012). In essence, there are informal or lightweight ontologies on one end, primarily geared towards some sort of communication and on the other end, formal ontologies help in automated reasoning of knowledge (Giunchiglia & Zaihrayeu, 2009). This paper falls in the middle and mostly uses OWL/XML Schemas to address the primary needs of communication and automation. They also provide a mechanism to use instructional design as a basis throughout the design of educational technologies. In addition, two key future research directions that motivate the need for ontologies are:

- Design of personalized learning environments for a diversified range of learners, teachers and subjects in India and across the globe
- Design of technologies that allow students to explicitly justify their answers through *reasoning* and provide a debugging environment through automated reasoning

Additionally, *scope* is another critical factor that can be used to classify ontologies at different levels of granularity. A distinction is made between *upper ontologies* that describe general-purpose concepts and their relationships, *domain ontologies* that define domain-specific concepts, *task ontologies* that specify domain-specific activities and *application ontologies* that instantiate domain ontologies and integrate task ontologies for a particular application (Guarino, 1998).

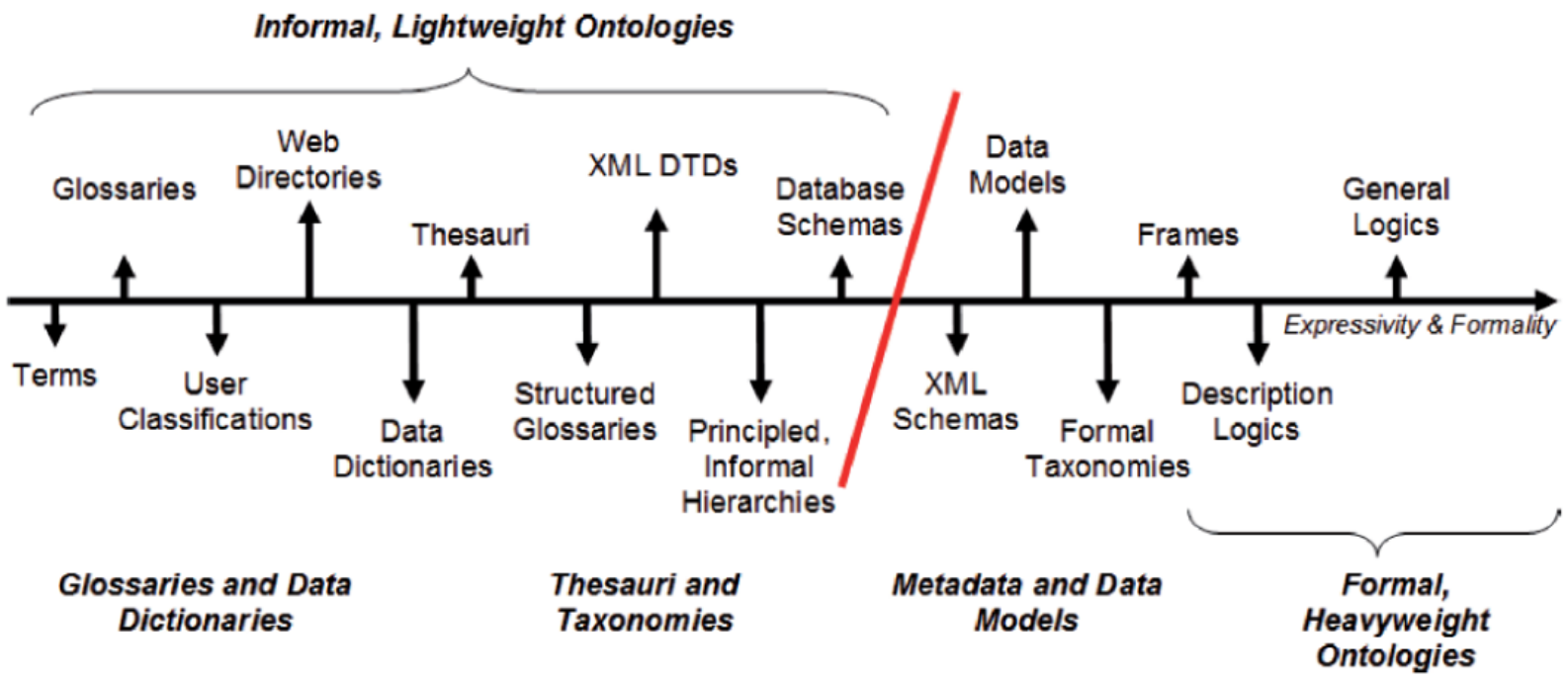

**Fig. 3** A spectrum of ontologies

### 3.2 Ontologies for Adult Literacy Instructional Design

In this section, we discuss our notion of ontology for adult literacy instructional design in relation to existing literature. The term ontology is used in a variety of ways in the literature (Uschold & Gruninger, 1996)(Studer, Benjamins, & Fensel, 1998)(Wong et al., 2012). A commonly used definition of ontology in computer science comes from Gruber (Gruber, 1993), where he defines an ontology as *"a formal, explicit specification of a shared conceptualization"*. This definition was further characterized and explained by several researchers (Uschold & Gruninger, 1996) (Studer et al., 1998) as:

– *formal*, the ontology should be represented using a formal language processable by machines and tools

  *This characteristic will allow us to represent instructional design knowledge in machine-processable form to facilitate automation*
– *explicit*, by using different types of primitives and precisely stating different concepts and axioms defining the ontology

  *Instructional design for adult literacy is embedded in IPCL and primers; ontologies will help in making it explicit*
– *shared*, the ontology is meant for a group of stakeholders within a community belonging to a specific domain or sub-domain

*How to share IPCL across all Indian languages and How to specify this knowledge to software engineers?*

- *conceptualization*, represents a specific view of a domain through various abstractions

*varied views of instructional designs for multiple contexts of adult literacy in India*

To the best of our knowledge, we could not find any ontologies that are even remotely connected to adult literacy in India. But we searched the literature for various ontologies focusing on different kinds of educational knowledge and give a few examples here. An ontology for literacy was proposed in the context of intelligent tutoring systems way back in 1999 (Carvalho & Pain, 1999). We then looked into some upper ontologies and found an example curriculum ontology devised by BBC for the national curricula on UK focusing on three topics (Algebra, Geometry, Formula), level (KS1, KS2, KS3, GCSE) and different fields of study (Maths, English, Science) (BBC, 2016). A comprehensive ontology that models several learning theories is presented in (Bourdeau, Mizoguchi, Hayashi, Psyche, & Nkambou, 2007) where the idea is to have solid pedagogical basis for intelligent tutoring systems. Heiyanthuduwage et al. have analyzed 14 ontologies developed by different institutions for learning design and proposed an OWL-2 learners profile (Heiyanthuduwage, Schwitter, & Orgun, 2016). One of the earliest ontologies developed by Mizoguchi focuses on creating a task ontology to facilitate reuse of problem solving knowledge (Mizoguchi, Vanwelkenhuysen, & Ikeda, 1995). We came across several ontologies focusing on particular kind of instructional design; for example, a mobile learning ontology was designed for abductive science inquiry style of instruction (Ahmed & Parsons, 2013). An ontology for learning scenarios based on collaborative learning theories in (Isotani et al., 2013) and one focusing on gamification is presented in (Challco, Moreira, Mizoguchi, & Isotani, 2014). There were other set of ontologies focusing on specific subject matter, like word problems in mathematics (Lalingkar, Ramanathan, & Ramani, 2015), software engineering body of knowledge (Abran et al., 2006). In addition to these kinds of ontologies, there are different kinds of ontologies developed for learning content (Verbert et al., 2004), learning design based on IMS LD standard (Amorim et al., 2006), a context ontology for bridging the gap between learning content and learning design (Jovanović, Knight, Gašević, & Richards, 2006). There were ontologies to represent learning object (Wang & Koohang, 2009) and learning design repositories (Paquette, 2014) to facilitate search and retrieval of learning resources on the web.

However, none of these ontologies cater to the need of *scale* and *variety* inherent in the problem domain and are not driven by patterns motivating the need for our proposed work. In the next section, we will briefly discuss the approach for development of ontologies.

## 4 Development of Ontologies

Ontology development has matured in the last few decades from being a research topic to even the discipline of ontology engineering with several approaches (Gomez-Perez et al., 2006)(Fernández-López & Gómez-Pérez, 2002). Specifically, ontology development methods supported with tools became quite popular in the recent times and *protégé* is one of the exemplar examples to illustrate this case. In their Ontology Development 101, Noy and McGuinness proposed an iterative approach for building ontologies consisting of several activities that need not be sequential (i) determine scope (ii) consider reuse (iii) enumerate terms (iv) define classes (iv) define properties (v) define constraints (vi) create instances. An important conclusion from their work is *"there is no single correct ontology for any domain. Ontology design is a creative process and no two ontologies designed by different people would be the same"* (N. Noy, McGuinness, et al., 2001).

We use OWL 2, a W3C recommendation that refines and extends OWL, the Web Ontology Language for representing knowledge in the semantic web (Group et al., 2009). OWL 2 is based on earlier version of OWL, extends RDF and is also compatible with XML. According to Krotzsch, OWL serves as a descriptive language for expressing expert knowledge in a formal way and as a logical language for drawing conclusions from that knowledge (Krötzsch, 2012). Accordingly, OWL 2 allows ontology engineers to represent knowledge using various representations like RDF/XML, OWL 2 XML, Functional Syntax, Macnhester Syntax, Turtle as shown in Table 1 with each of the methods having different expressive power and reasoning abilities (Group et al., 2009). The choice of ontology representation is primarily decided by ontology engineers based on the requirements and needs of the domain (Guarino, 1998). We confine ourselves to the descriptive use of ontologies and use OWL/XML for representing ontologies in this paper.

**Table 1** Syntax variations in OWL 2 Web Ontology Language

| **Name of Syntax** | **Specification** | **Purpose** |
|---|---|---|
| RDF/XML | Mapping to RDF Graphs | Interchange (can be written and read by all conformant OWL 2 software) |
| OWL/XML | XML Serialization | Easier to process using XML tools |
| Functional Syntax | Structural Specification | Easier to see the formal structure of ontologies |
| Manchester Syntax | XML Serialization | Easier to read/write DL Ontologies |
| Turtle | Mapping to RDF Graphs, Turtle | Easier to read/write RDF triples |

We see three major directions for developing ontologies from a synthesis of the literature (i) manually by expert(s) for specific purposes following a varied set of processes from light-weight to a rigorous standard process (ii) semi-automatic way of developing ontologies, where a part of the ontology is developed manually and a part is automatically retrieved using text mining, natural language processing and other machine learning techniques (iii) fully automatic development, where the ontologies are derived using ontology learning approaches.

We follow a simple process for developing ontologies in this paper based on existing literature. The first step in the process is to determine the requirements from the ontology, which is driven by the set of *e*Learning Systems to be developed in our case. The next step is to figure out the scope of the ontology drawing a boundary for what is within and outside the scope. Once the scope is defined, the next step is to identify any existing ontologies that can be used for creating the ontology. There are several search engines like SWOOGLE[6] and ontology servers like OntoLingua[7]s for searching existing ontologies.

We discuss the ontologies we adapted from existing literature in the next section. Once the suitable sources for ontologies are defined, the next step is to use a standard approach to identify the concepts, relationships between the concepts, define properties, constraints and instances using an appropriate representation language like OWL/RDF. An important distinction of this process from the standard ontology development methodologies is the use of patterns as one of the critical sources for building ontologies. The patterns themselves are discovered after extensive discussions with domain experts; rigorous analysis of literature and analyzing existing applications that are built in the domain. We have extensively discussed with domain experts from NLMA; analyzed literature on adult literacy and instructional design as a source for our patterns. We also analyzed several *e*Learning Systems developed by TCS for 9 Indian languages before creating the patterns. We use these patterns as one of the primary source for creating the ontologies. We also consider other literature from the instructional design space as input to our ontologies. The output of this entire exercise of conceptualization and implementation is a set of ontologies. The evaluation of this ontologies is carried out by developing a set of applications based on these ontologies and assessing whether the domain requirements have been met or not. Figure 4 shows a part of how we devised scope for our instructional design ontology framework. We detail our framework in the next section.

## 5 IDont - An Ontology Based Framework for Modeling Instructional Design

Is teaching science the same as teaching mathematics? Does teaching in a country like US and India same? Can we use the same method for teaching different kinds of learners? The answer for most of these questions is no. There have been tremendous efforts in trying to come up with several standards in the space of educational technologies such as SCORM for learning objects (Bohl, Scheuhase, Sengler, & Winand, 2002), IMS-LD for learning designs (Koper, 2005), IEEE LOM for learning objects (Neven & Duval, 2002) and enormous research (Burgos, 2015), platforms and tools (Botturi, Stubbs, & Global, 2008) surrounding these standards. However, despite significant progress, most of the promises in educational technologies seem to be unfulfilled (Toyama, 2011). We summarize the following major pitfalls:

– One-size-fits-all - There are hundreds of learning theories in the literature. Attempts towards coming up with a unified way of dealing with them turned quite complex denting the success of the initiatives.
– *End-to-end automation* - Several attempts have been made to automate different aspects of education, ranging from modeling learning theories to automatically generating learning environments and this focus on end-to-end automation turned futile in the most of the attempts (Goddard, Griffiths, & Mi, 2015). For instance, in the case of IMS-LD, even though not stated explicitly, this goal of end-of-end automation resulted in complex authoring (Burgos, 2015).
– *Administration and Management* - While it is important to handle and ease the job of teachers in administration and management activities for which several Learning Management Systems were developed, linking the instructional design to LMS has increased further complexity with the existing approaches.

[6] http://swoogle.umbc.edu/

[7] http://www-ksl.stanford.edu/knowledge-sharing/ontolingua/

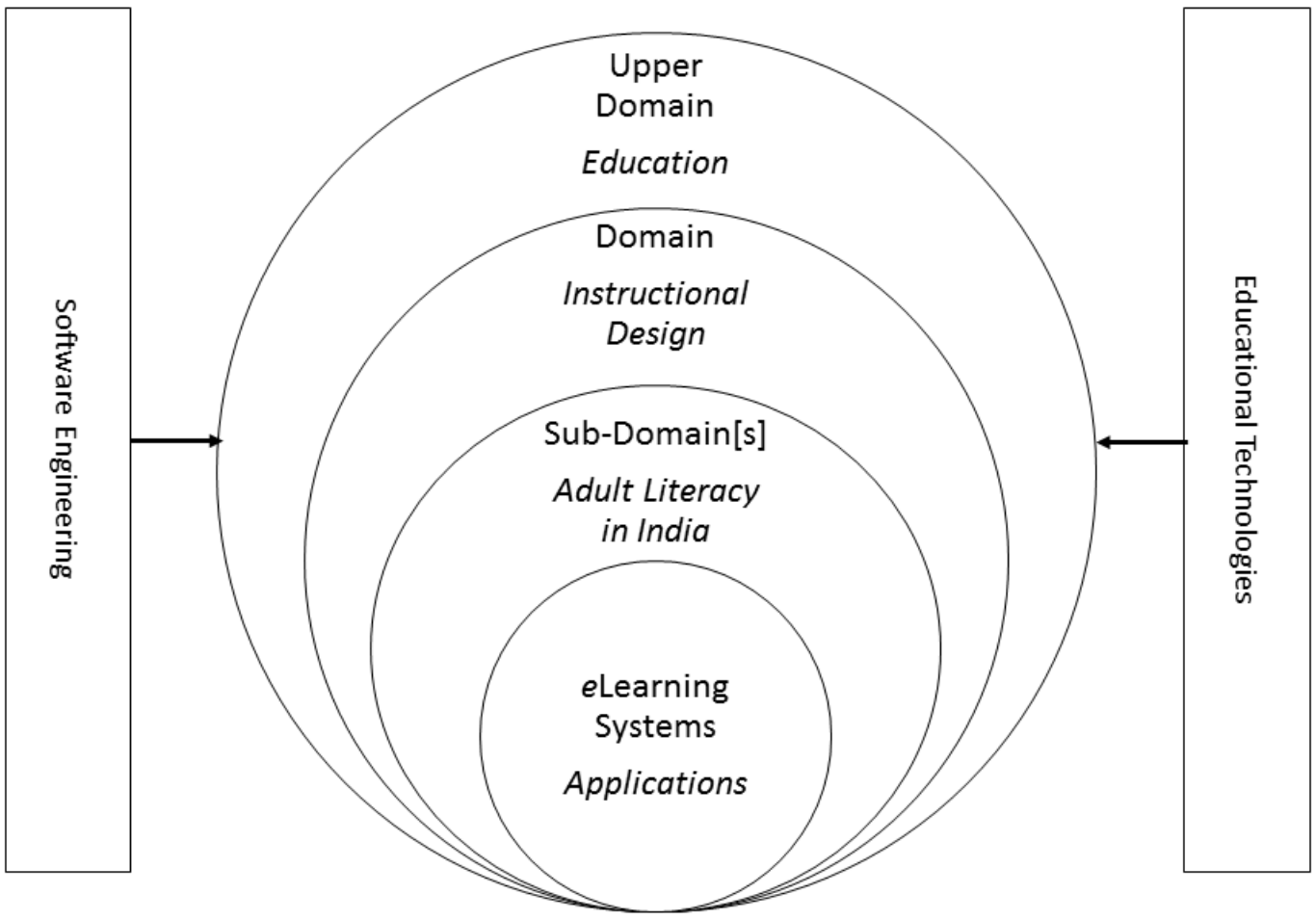

**Fig. 4** Scope of Ontologies in this paper

We learn from these experiences and propose a framework for modeling instructional design using *ontologies* based on *patterns*. The design rationale for ID*ont* is as follows:

– *Simplicity & Separation of concerns approach* - We strongly advocate the separation of concerns principle (Dijkstra, 1982) to model ontologies. The core idea is to have smaller multiple ontologies for different aspects of instructional design such that they can be adapted, extended and reused in other contexts. Figure 5 shows how different aspects can be separated as components having explicit interfaces such that they can be connected with other aspects and customized for a specific learning situation.
– *Leverage and Reuse existing ontologies* - When designing new ontologies, ontological engineering suggests the utilization, adaption and extension of existing ontologies (learning objects, learning designs).
– *Technology Design* - The framework should support the creation of a platform and authoring tools to explicitly capture and model all the ontologies.
– *Extensibility and Customization* - These are two major criteria for the framework as most of the times the ontologies have to be customized and extended for the specific domain. There should be a provision in the framework such that existing ontologies should be replaced with custom ontologies with minimal effect on the overall framework.
– *Iterative and Collaborative Approach* - The design of this framework should follow an iterative approach and must consult different stakeholders (such as teachers, learners, instructional designers and so on) during the process.
– *Internationalization* is required for both ontologies and tools that support the creation of ontology instances.

The core premise of this framework is to systematically model instructional design using different aspects like *context*, *goals*, *process*, *content*, *evaluation*, *environment*. We distinguish between two kinds of instructional design knowledge, one is at a *conceptual* level that maps with existing learning methodologies and the other is at a *technical* level to facilitate semi-automation of *e*Learning Systems. In this context, the core idea of ID*ont* is not to define complete ontologies but to point to several possible modular ontologies that are required for systematic modeling of instructional design. As such, most of the aspects of ID*ont* are optional and can be configured based on specific purposes and learning situations. Figure 5 shows an overview of the ID*ont* framework. The key inputs come from a set of instructional design requirements that drive selection of appropriate instructional design models, which are captured as patterns in our approach. We do not specify the exact ontologies for instructional design but have placeholders for different aspects. With the advent of several ontology repositories, an instructional designer or ontology engineer can either extract the required ontologies for specific instructional design model from existing literature or create a new one. This generic instructional design stitched from existing or new ontologies can be customized with domain ontologies and is further realized by specific instances like ID1, ID2 ... ID*n*.

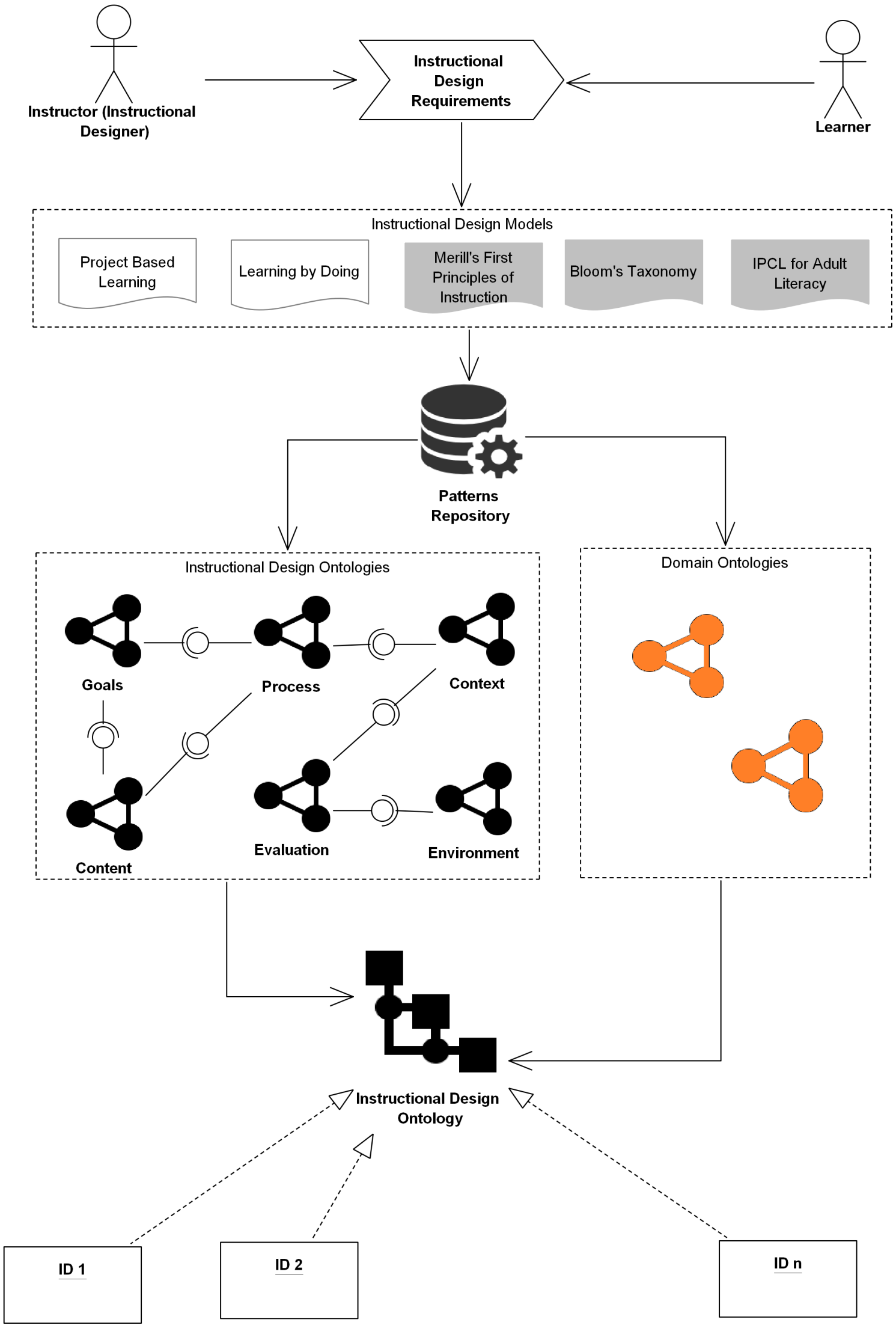

**Fig. 5** Overall Process of IDont Framework

Figure 6 presents an overview of ID*ont* framework for adult literacy. Even though we show several ontologies in the diagram, we focus our discussion on *goals*, *process* and *content* ontologies. We briefly explain the important ontologies of our framework as follows:

**A. ContextOntology** - Context plays a significant role in ID*ont* as it allows for modeling of various aspects related to a particular learning situation. The notion of learning context was proposed in LOCO ontology to bridge the gap between learning design and content consisting of domain specific information (Knight et al., 2006). However, in this framework, we articulate context in a broader view that encompasses several pointers to all other ontologies. This is a meta-ontology that essentially captures the basic information related to all other aspects of instructional design such that each of these aspects can be potentially (re)used. As shown in Figure 6, *ContextOntology* has metadata associated with it along with context information related to various aspects of instructional design. *ContextOntology* specifies how a *ProcessContext* achieves *goals* using *ContentContext* delivered through *EnvrionmentContext* following *EvaluationContext* and performed by *RolesContext*.

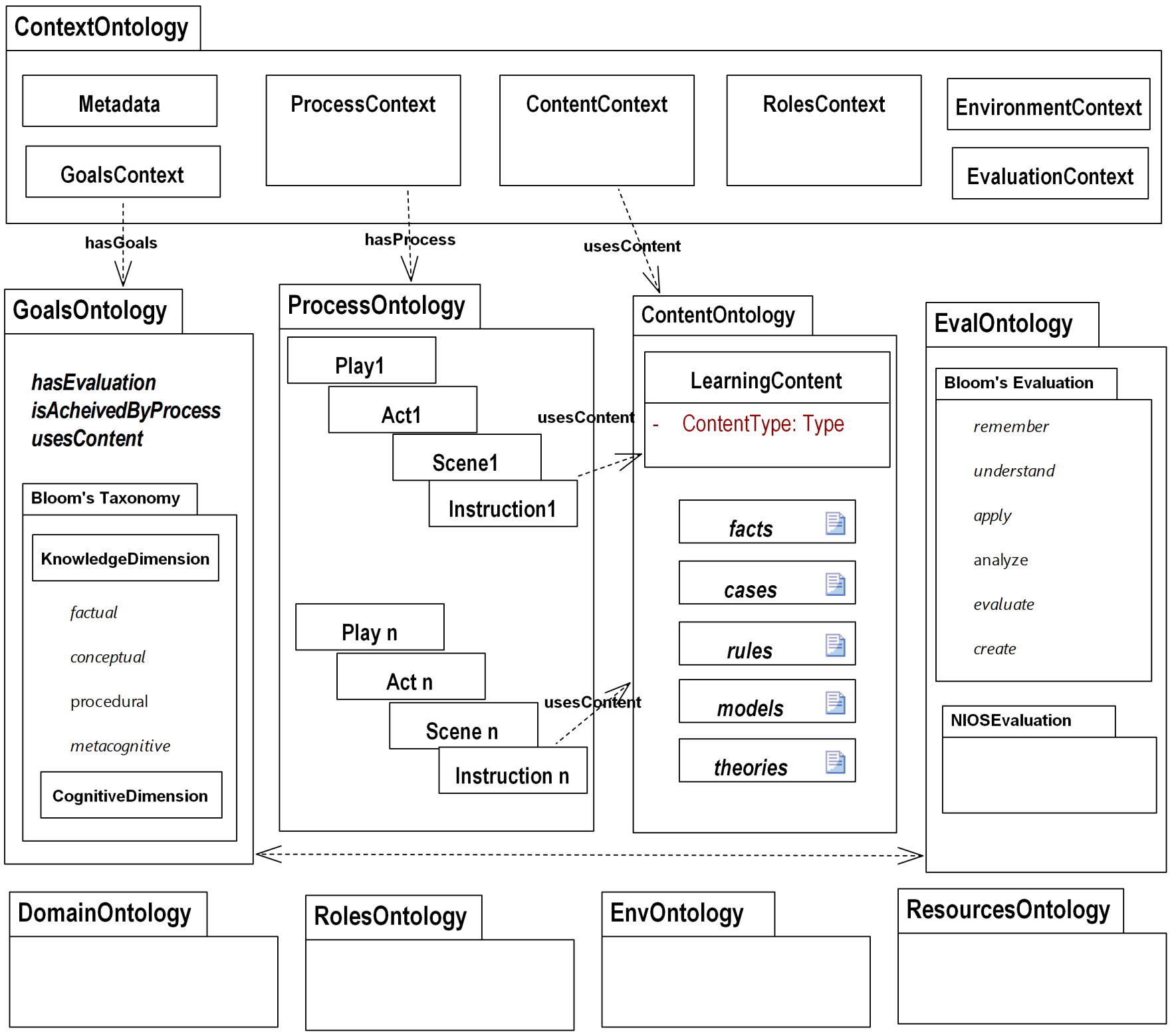

**Fig. 6** Instructional Design Ontology for Adult Literacy

**B. GoalsOntology** - This ontology formalizes the notion of goals (which can be instructional goals, learning goals or even learning outcomes). The details of how it is defined are left to the specific instance. Some properties associated with goals are *hasName*, *hasPriority*, *hasPrerequisites*, *hasEvaluation*, *isAchievedByProcess*. The *GoalsOntology* points to the process through which these goals will be achieved, target competencies, the instructional material that is required and the evaluation to be performed. Consider the scenario of creating goals for K12 students, and goals are prescribed by education boards.

Teachers can potentially reuse these goals if captured in the form of *GoalsOntology*. As the evaluation related to these goals is separately captured, it can be reused as well. We prescribe to the idea of *goal-driven instruction* as part of our framework irrespective of instructional design models. We detail *GoalsOntology* in the latter part of this paper in Section 5.1.

**C. RolesOntology** - The success of instructional design depends on several people who perform their roles in the process. Hence, it is important to capture the roles, their responsibilities and how the different aspects of instructional design should be adapted according to the needs of specific roles. The two most important roles are that of teacher and learner. The crucial knowledge about a learner is captured through learner profiles consisting of several attributes. Generic roles in instructional design are captured first and then modified based on specific learning situations. Mapping of goals with competencies of roles can also be done in this ontology. Most of the roles are further associated with teams and in that case the role of the team as well as individual persons is modeled separately along with roles performed by agents. This ontology should also have metrics to trace from goals to evaluation.

**D. ProcessOntology** - The crux of ID*ont* framework is the *ProcessOntology* that captures the instructional design process, and relates to all other ontologies and practically executes the process. In the literature, Learning Design is discussed heavily, in particular IMS Learning Design (Consortium et al., 2003) and received criticism as well (Burgos, 2015). Ontologies for modeling learning design are presented in (Amorim et al., 2006). Based on our prior experience with adult literacy instructional design, IPCL and our future goal to introduce reasoning into educational technologies, we proposed the *ProcessPattern* - (*play*, *act*, *scene* and *instruction*) (Chimalakonda & Nori, 2014). Each lesson is organized as hierarchy of *pasi* with instructions where concrete activities are performed based on Merill's principles of instruction in this particular instance. This instruction actually points to *ContentOntology* and associates required content for the respective instruction. This nomenclature allows us to systematically capture the knowledge of instructional design process and potentially reduce technological effort. This hierarchy has similarities to IMS LD but has variations to align with *patterns* for adult literacy instruction. We will present the *ProcessOntology* in Section 5.2.

**E. ContentOntology** - This ontology allows for modeling of instructional material in a particular learning situation. There is extensive research on ontologies for learning objects and we use the ALOCoM ontology (Verbert, Jovanovic, Duval, Gasevic, & Meire, 2006) as base for content aspect of our framework. However, for adult literacy instructional design, we have used *fcrmt* (facts, cases, rules, models, and theories) structure (Chimalakonda, 2010) as discussed in Section 5.3. So the *ContentType* of ALOCoM also includes *fcrmt* to support reasoning. The *ContentOntology* is closely associated with other ontologies and strongly with the *ProcessOntology*.

**F. EvaluationOntology** - What if the most common evaluations of instructional design are captured and an instructor can customize them based on his or her requirements? The main intent of this ontology is to capture evaluations as independent knowledge and link them with goals through *ContextOntology*. This separation makes it easier to perform different kinds of evaluations for the same set of goals. This ontology captures the details of evaluation and has a direct relationship with *GoalsOntology* which is connected with *ProcessOntology*.

**G. EnvironmentOntology -** The final execution of instructional design happens in an environment. Consider the scenario of a lesson to teach about shapes to K2 students. Considering that the students first listen about shapes in a *ClassroomEnvrionment* and if a teacher wants to use computers, then most of the other aspects of instructional design remain same except the environment which changes to *ComputerEnvironment*. Separating the environment from the rest of instructional design makes it easier to run the learning situation in different environments similar in spirit with software deployment.

**H. DomainOntology -** This ontology mainly articulates and customizes key aspects of instructional design with respect to a specific domain and provides a domain-specific version of ontology. In particular, the various sub-ontologies and properties of these ontologies will have detailed associations when mapped to a specific domain. For example, the *ContentOntology* will have strong co-relation and mapping with content in the domain. We present a domain-specific ontology for adult literacy in 6.1.

There can be several other ontologies like *ActivitiesOntology* for capuring various activites that are suported in the instructional design, *WorkflowOntology* to model the tedious workflows in education, *FeedbackOntology* to capture continuous feedback of the instructional design, *OrganizationOntology* focusing on characteristics of the organization, *ResourcesOntology*, having pointers to specific resources like text, audio, video and so on. In our analysis of instructional design literature, we strongly see that it is virtually impossible to capture all kinds of instructional design models and any attempt towards it turns to be futile. However, the main intent of our framework is to use a separation of concerns approach to systematically capture various aspects of instructional design through ontologies.

We discuss the specific ontologies that are developed as part of ID*ont* framework. Our attempt is not to present complete ontologies but to design educational technologies in sync with instructional designs and for scale and variety. We also include several entities in the ontologies for future use rather than just current needs. We also rely on the practice of adapting existing ontologies and create new entities only if required (N. Noy et al., 2001). We discuss some core ontologies of our framework in the coming sections and illustrate them through adult literacy case study.

### 5.1 An ontology for modeling goals

The primary goal of any instructional design is to find ways to support learners in achieving their learning goals (Ram & Leake, 1995). Based on the pattern discussed for goals in (Chimalakonda, 2017), we present an ontology for representing instructional goals in this section based on revised Bloom's taxonomy (Anderson et al., 2001). Figure 7 shows a part of *GoalsOntology*[8] developed using *protégé*[9] tool from Stanford. The priority of the goal is described using *GoalPriority*, progress through *GoalProgress*, deadline through the property *goalDeadline*. An important sub-class is to classify the goal according to a taxonomy. The class *GoalClassification* is further divided into two classes *BloomTaxonomy* and *ABCD*.

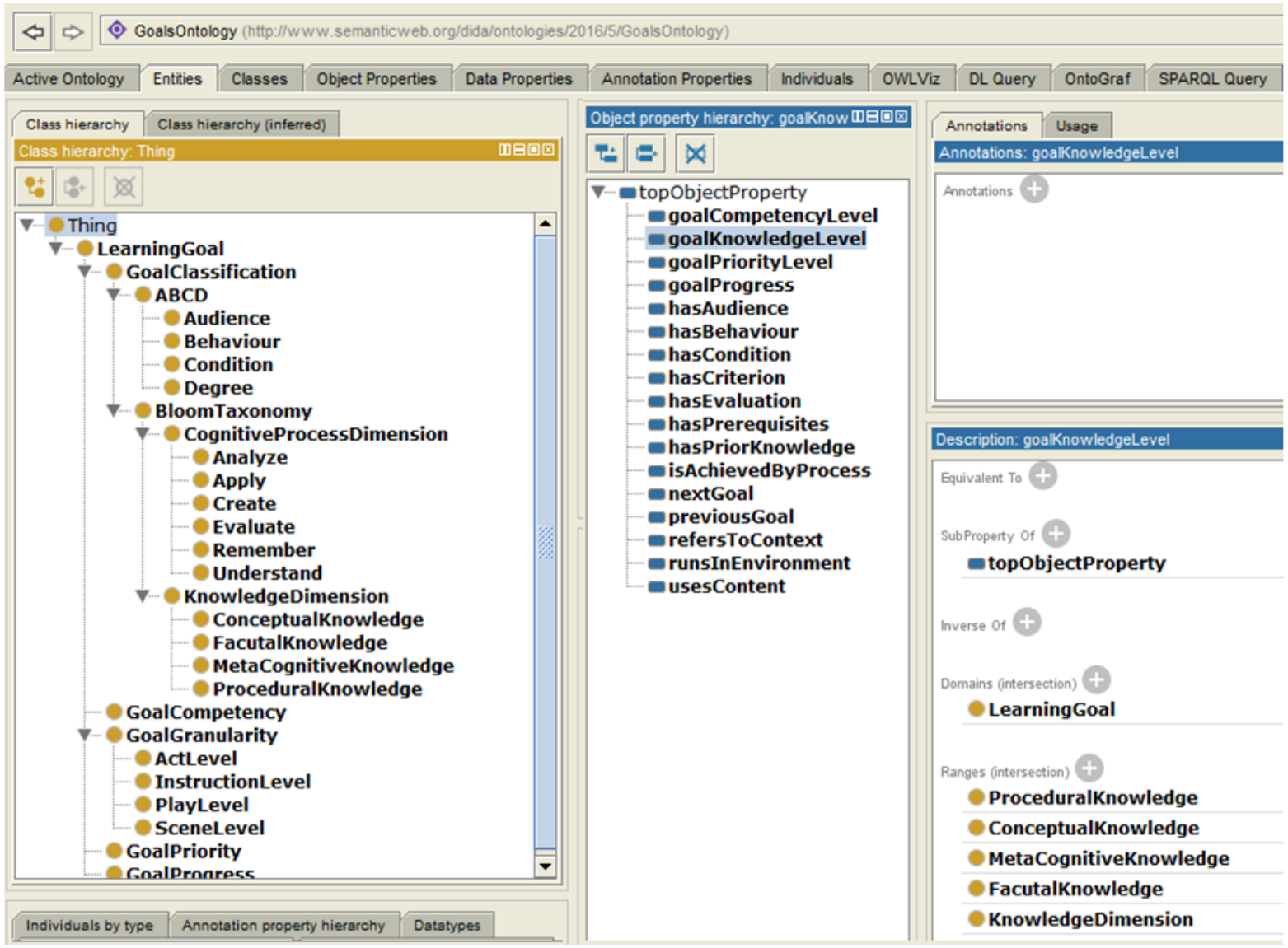

**Fig. 7** A fragment of *GoalsOntology*

The *BloomTaxonomy* is further divided into *KnowledgeDimension* and *CognitiveProcessDimension* as per revised Bloom's taxonomy. The knowledge can be classified as *FacutalKnowledge*, *ConceptualKnowledge*, *ProceduralKnowledge* and *MetaCognitiveLevelKnowledge* with increasing levels of higher order levels of thinking. This is in sync with the *ContentOntology* that will be discussed in next section 5.3. The *CognitiveProcessDimension* is the most commonly used way to classify goals as per Bloom's taxonomy. It has six levels *Remember*, *Understand*, *Apply*, *Analyze*, *Evaluate*, *Create* and each of them have several verbs specifying the activities.

Several object properties are shown in Figure 7 connecting different concepts in the ontology. Priority of the goal can be captured using *goalPriorityLevel*, competency through *goalCompetencyLevel* and *goalKnowledgeLevel* can have a range of values from the *KnowledgeDimension* and maps to the *fcrmt* pattern. Every goal should have a *goalDeadline* and its progress is monitered through *goalProgress*. A goal also has *hasPrerequisites*, *previousGoal* and *nextGoal*. This ontology is connected to *ProcessOntology* through

[8] A detailed overview of this ontology is available at https://goo.gl/wdRU5b.
[9] http://protege.stanford.edu/

*isAchievedByProcess*, *ContentOntology* via *usesContent*, *EvaluationOntology* through *hasEvaluation* and *runsInEnvironment*.

In addition, there are several data properties that are associated with the ontology. For example, *goalDeadline* stores the deadline as *dateTime*. The goal itself can be described using *goalText*, *goalImage*, *goalAudio*, *goalVideo*, *goalMetadata*. These data properties store specific information that can be later used for (semi-)automatic generation. *GoalGranularity* is another critical class that is specific to our instructional design as we have a goals hierarchy akin to *play*, *act*, *scene*, *instruction* pattern. In addition to the standard concepts, the ontology also has concepts for *GoalPattern* consisting of properties shown in Figure 7. For example, *SourceOfPattern* is a data property that specifies the source of the patterns, *Trade-Offs* specifies the issues that might occur using this pattern. In our case, we realized that if specifying *goals* requires so much of effort, the entire exercise will be a burden for teachers and instructional designers making it a futile effort in the end. Hence, we have minimal mandatory properties with scope for using extended properties only if required.

### 5.2 An ontology for modeling instructional processes

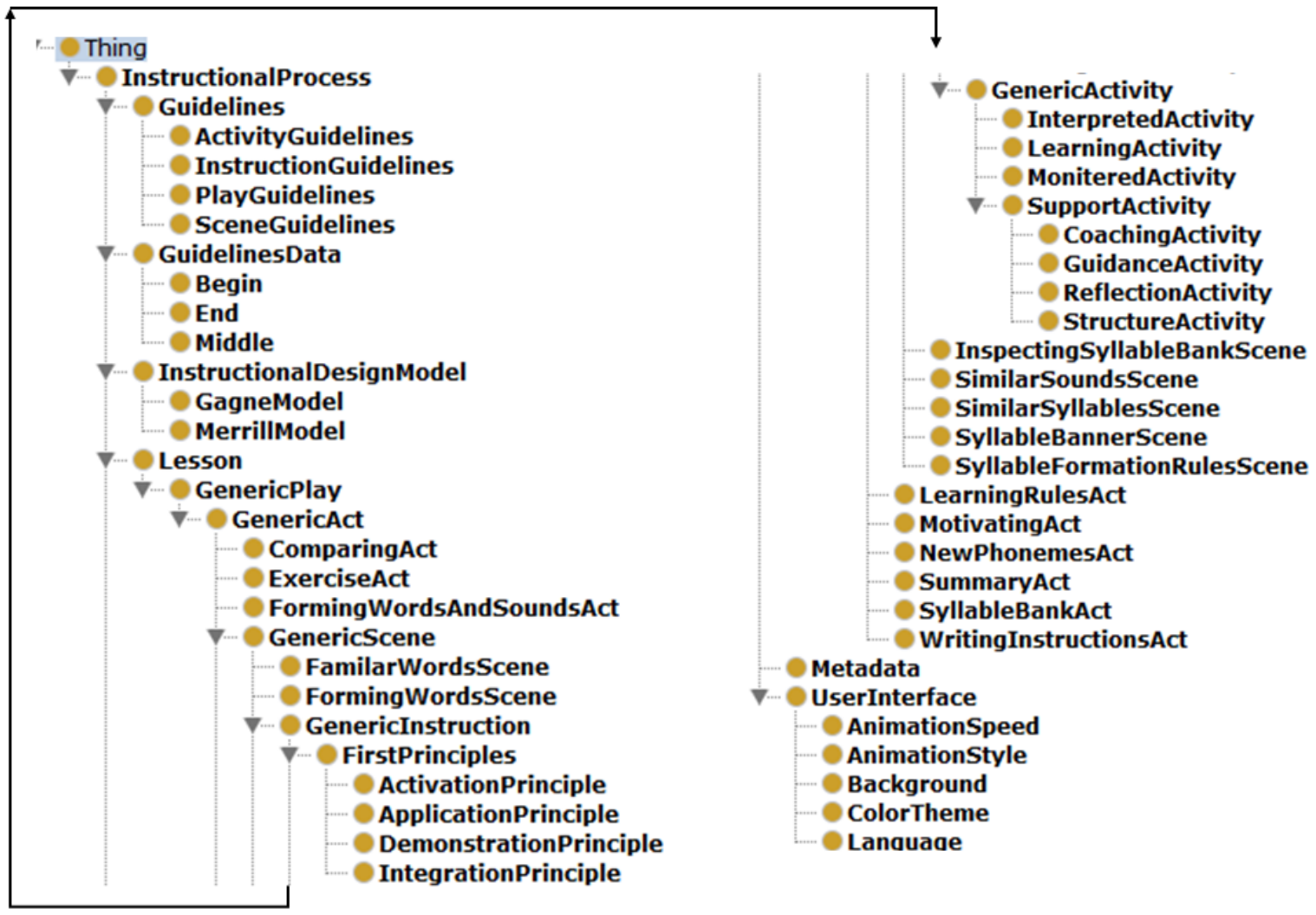

**Fig. 8** A fragment of *ProcessOntology*

The *ProcessOntology*[10] is a core ontology for specifying instructional process and is closely associated with several other ontologies. As shown in Figure 8, the ontology is divided into three conceptual sections at a higher level (i)

[10] A detailed overview of this ontology is provided in https://goo.gl/5A937v.

*learning*, focusing on concepts that map to the underlying learning methodologies (ii) *metadata* consists of information about the process in general (iii) *user interface* to declaratively specify a few aspects of the *e*Learning System. The *ProcessOntology* is based on *ProcessPattern* (Chimalakonda & Nori, 2014)(Chimalakonda, 2017)and its primary purpose is to achieve goals specified in the *GoalsOntology* and is connected through *hasAssociatedGoal* property. These goals have to be achieved using content specified via *ContentOntology* connected through the object property *usesContent*. Similarly, *usesEvaluation*, *performedbyRole*, *runsInEnvironment* connect this ontology to *EvaluationOntology*, *RolesOntology* and *EnvironmentOntology* respectively. This ontology has several data properties like *title*, *description*, *metadata*, *noOfPlays*, *noOfScenes*, *noOfInstructions*. One important property is *hasTimeLimit* that specifies the time limit for an *activity*, *instruction*, *scene*, *act*, *play*. *Guidelines* is an important concept that we use for giving instructions to learners during their interaction with the *e*Learning System at different levels of granularity specified using *PlayGuidelines*, *ActGuidelines*, *SceneGuidelines*, *InstructionGuidelines*, *ActivityGuidelines*. For example, a guideline from a teacher might be *"Everybody look at the screen and observe how the two syllables are combined together to form a new word"*. Separating this information provides the flexibility to change guidelines. This can be specifically used to change *medium of instruction* in an *e*Learning System. A language like *Hindi* can be taught using *Telugu* as medium of instruction by changing the guidelines in the entire system.

The base *InstructionalDesignModel* can be specified as *MerrillModel* or any other instructional design model from the literature. We use *MerrillModel* as it is based on first principles of instruction distilled from several instructional designs (Merrill, 2012). Then each *lesson* is modeled using a set of plays (*GenericPlay*) that are divided into acts (*GenericAct*), which are further divided into scenes (*GenericScene*) and instructions (*GenericInstruction*). We have identified different kinds of acts for adult literacy instruction which include *MotivatingAct*, *NewPhonemesAct*, *FormingWordsAndSoundsAct*, *SyllableBankAct*, *ComparingAct*, *LearningRulesAct*, *WritingInstructionsAct*, *ExerciseAct*, *SummaryAct*. We inferred these acts from adult literacy eLearning Systems that are tested on the field. There are different kinds of scenes *SimilarSoundsScene*, *SimilarSyllablesScene*, *InspectingSyllableBankScene*, *SyllableFormationRulesScene*, *FamilarWordsScene*, *SyllableBannerScene*, *FormingWordsScene* under each act. Each scene further has instructions which have direct activities for facilitating learning. Each instruction follows one or more principles and can have one or more activities. We specify Merrill's first principles of instruction using *FirstPrinciples* that is further divided into *IntegrationPrinciple*, *ActivationPrinciple*, *DemonstrationPrinciple*, *ApplicationPrinciple*, *DemonstrationPrinciple*. Activity is one of the most commonly used concept in the space of instructional design and we model that using *GenericActivity*. We incorporate two kinds of activities from the literature *LearningActivity* and *SupportActivity*. But we also model four kinds of additional activities *StructureActivity*, *GuidanceActivity*, *CoachingActivity* and *ReflectionActivity* to accommodate Merrill's inner circle of structure-guidance-coaching-reflection. Modeling these activities as concepts allows us to change these activities based on learner styles or instruction styles. In addition, *InterpretedActivity* and *MoniteredActivity* help from evaluation perspective.

The current ontology also has basic concepts for *UserInterface* such as *AnimationStyle*, *ColorTheme*, *AnimationSpeed*, *Language*, *Background*. The instances of these concepts will help in configuring the user interface of *e*Learning Systems for adult literacy based on specific requirements.

One principle behind this ontology is not to use all the classes and properties but to further filter this ontology to the specific needs and use only a fragment of the ontology in order to reduce the burden on the teachers and instructional designers. For example, if a course has 1000 instructions in total, then specifying principles for all of these instructions might be a burden and an alternative could be to make this property optional at instructional level but mandatory at a *scene* or *act* or *play* level.

### 5.3 An ontology for modeling instructional material

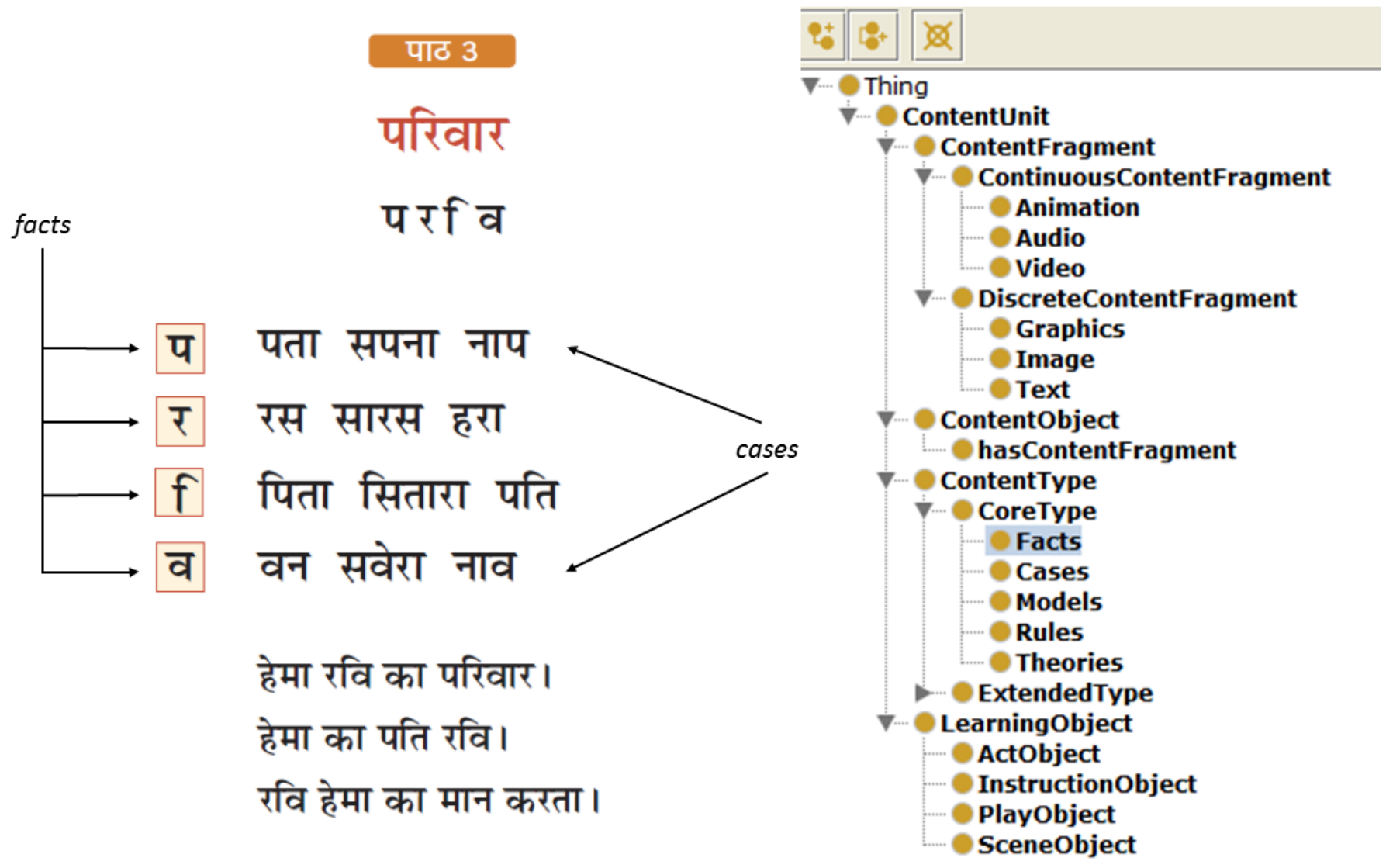

**Fig. 9** A fragment of *ContentOntology*

The *ContentOntology*[11] ontology is primarily derived from existing literature on learning objects and specifically the ALOCoM ontology (Verbert et al., 2004) along with the *ContentPattern* elaborated in (Chimalakonda, 2017). As shown in Figure 9, this ontology includes four core concepts *ContentType*, *ContentFragment*, *ContentObject*, *LearningObject*. The raw data in the form of *text*, *audio*, *animation*, *video* are concepts in *ContentFragment* and *ContentObject* is an aggregation of several content fragments. This ontology is further refined in terms of *ContentType*, which includes *Facts*, *Cases*, *Rules*, *Models*, *Theories*, which form the *CoreType*. In *ExtendedType*, there are further concepts derived from the literature (Verbert et al., 2004). Essentially, they capture learning objects at a higher level of abstraction. Another important concept is *LearningObject* which has the sub concepts of *PlayObject*, *ActObject*, *SceneObject*, *InstructionObject*. These concepts are connected to respective elements in *ProcessOntology*.

[11] A detailed overview of this ontology is provided in https://goo.gl/ZSEo5a

There are other ontologies for specifying *Roles*, *Evaluation*, *Environment* that are part of instructional design ontology but defining those ontologies is beyond the scope of this work. The *RolesOntology* is an interesting one with roles like Teacher, Mentor, TeachingAssistant, Coach and so on and can be mapped to different kinds of activities in the *ProcessOntology*. As an example, learning styles and teaching styles may be used in the role of learner and teacher in *RolesOntology*. In the next section, we will briefly discuss a domain ontology for adult literacy.

## 6 Evaluation

There exist several ways of evaluating ontologies in the literature (Brank, Grobelnik, & Mladenic, 2005). Broadly, these approaches can be classified as (i) manual, mainly driven by human interventions, either experts or users (ii) automated approaches and (iii) semi-automated approaches that fall in between. To reiterate, the goal of this paper is to facilitate scale and variety during design of educational technologies rather than improving quality of instruction through knowledge representation using ontologies. To this end, we presented an ontology based modeling framework in this paper rather than an ontology. As part of our evaluation, (i) we demonstrated our framework through adult literacy case study in detail throughout the paper (ii) we also present a domain ontology that is an instantiation of the proposed framework for adult literacy in this section (iii) the ontology framework was used as a base for creation of a platform `http://rice.iiit.ac.in/al.html` for generating *e*Learning Systems for adult literacy in India. This platform was further used to generate 9 *e*Learning Systems (Chimalakonda, 2017) but in principle can be used to create thousands of *e*Learning Systems.

### 6.1 A Domain Ontology for Adult Literacy

A distinction has been made in the literature between domain and application ontologies (Guarino, 1998). *Domain ontologies* are aimed towards defining concepts pertaining to a specific domain like *"adult literacy for Indian languages"* whereas *application ontologies* are further refined to specific needs of an application, in our case it can be an *e*Learning System for a particular language. According to Census 2011, there are about 29 languages spoken by more than a million people, 60 languages by more than 100,000 people and about 122 languages by more than 10,000 people speaking it in India. Of these, there are 22 official languages. A fundamental tenet of Indian languages is that they share a common phonetic base (Mudur, Nayak, Shanbhag, & Joshi, 1999). This commonality across a family of Indian languages provides a shared domain that can be used for representing domain knowledge as an ontology. Based on this premise, NLMA has come up with IPCL, a uniform methodology as explained in Section **??**, which is the base for creating instructional material for all Indian languages. In Indian languages, the term "aksharas" is used to refer to *alphabet*. This *akshara* refers to a sound that is formed of sounds of vowels and consonants (Singh, 2006). Being invariant of position is an interesting characteristic of *akshara* that holds for all Indian languages (Singh, 2006).

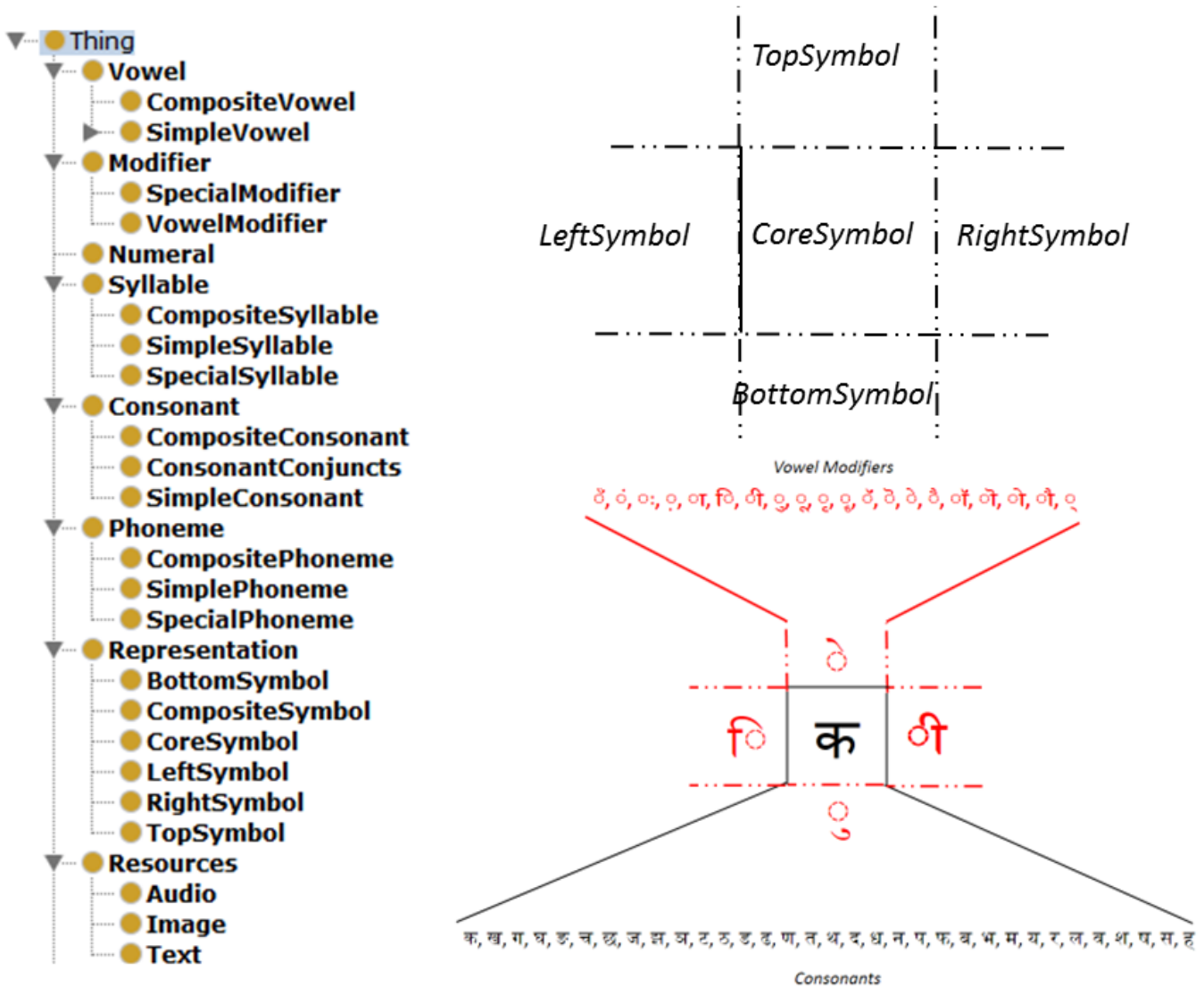

**Fig. 10** A fragment of *DomainOntology* for adult literacy

Figure 10[left] shows an ontology of Indic scripts for literacy, primarily focusing on the structure of syllables in the language. The core concepts of the ontology include *Syllable* denoting the visual representation of an *akshara* or a fragment of it. This is further specialized into *SimpleSyllable*, *CompositeSyllable* and *SpecialSyllable*. Every concept in this ontology have two properties *hasRepresentation* and *hasResources* refering to *Representation* and *Resources* respectively. *Representation* has five other concepts to systematically capture a syllable. Figure 10[right] shows the core structure of a syllable in Indian languages. The syllable itself is composed of *CoreSymbol*, *LeftSymbol*, *RightSymbol*, *TopSymbol*, *BottomSymbol*. Each of these concepts store the respective visual fragments of the syllable at the relative positions. For example, a base consonant like क can be modified using any of the vowel modifiers from left ि to give कि, adding the right symbol ी to क results in की. Similarly, adding

top symbol ◌े gives के and bottom symbol ◌ु results in कु. The composition of these symbols is represented as *CompositeSymbol*. The most important property of Indian languages is the broad choice of each of these symbols giving rise to an entire alphabet for a particular language. Figure 10[right] shows how vowel modifiers can be applied on four sides of a base consonant to give a set of composite symbols in a language. For most of the Indian languages, the number of vowel modifiers is 12 eventhough there are a few languages where the number can be less. Similarly, the number of consonants will vary from 12 to 36 for different languages.

The concept of *Resources* in the ontology helps in storing the data for respective syllables and phonemes in the form of *Text*, *Audio*, *Image*. We have observed that most of the eLearning Systems for Indian languages developed today rely on images for storing and displaying language aspects making it difficult to change the system. However, in our ontology we make an attempt to systematically separate different aspects to facilitate variety for a multitude of languages. The ontology also has the concept of *Vowel*, that is further divided into *SimpleVowel* and *CompositeVowel*. They are bound to *Syllable* through the property *hasVowel*. Similarly, *Consonant* class represents the consonants in a given language and can be either *SimpleConsonant* or *CompositeConsonant*. Most of the Indian languages have *ConsonantConjuncts* like కృ,క్ఖ,క్గ,క్ఘ,క్జ in *Telugu* language, whcih are special symbols formed with a number of syllables. *Modifier* is the core class for representing different modifiers like *VowelModifier* and *SpecialModifier*. *VowelModifier* in general consists of several signs often called as dependent vowels ◌ँ, ◌ं, ◌ः, ◌़, ◌ा, ◌ि, ◌ी, ◌ु, ◌ॢ, ◌ृ, ◌ॄ, ◌ॅ, ◌ॆ, ◌े, ◌ै, ◌ॉ, ◌ॊ, ◌ो, ◌ौ, ◌्. In Devanagari, there are twelve signs which when composed with consonants give rise to a number of composite syllables and is often called *Barakhadi* because of 12 modifiers. In addition to these, there are several special modifiers specific to a language. We have modeled *Phonemes* in similar lines to *Syllables* through respective concepts. *Numeral* class denotes the representation of symbols ० १ २ ३ ४ ५ ६ ७ ८ ९ in a particular language. We did not get into writing part in detail even though we have left scope for further extension of the ontology. This entire exercise of creating detailed ontologies for adult literacy is to systematically specify different parts such that they become source of variety to facilitate semi-automatic development of *e*Learning Systems.

We summarize the conclusions and future directions of our work in the next section.

## 7 Conclusions & Future Work

Instructional Design is one of the fundamental pillars of educational technologies and forms the basis for rest of the activities that drive effective instruction. Berger defines instructional design as a *"systematic development of instructional specifications using learning and instructional theory to ensure the quality of instruction"* (Berger & Kam, 1996). There are over 100+ instructional design theories in the literature catering to a diversified range of needs in education domain (Reigeluth & Carr-Chelman, 2009). In this paper, we motivated the need for modeling instructional design knowledge through ontologies to address *scale* and *variety* inherent in the domain. The key premise of the research presented in this paper is to systematically model different aspects of instructional design as modular ontologies such that these modular ontologies can be composed together to represent an instructional design. We also showed how changing any of these ontologies will result in a variant of the instructional design. We specifically presented an ontology for modeling goals, an ontology for modeling instructional process and an ontology for modeling instructional material. We demonstrated each of these ontologies through adult literacy case study which requires thousands of similar but distinct *e*Learning Systems to be developed. The systems that are developed based on these ontologies are made available at http://rice.iiit.ac.in and transferred to National Literacy Mission of Government of India.

As pointed by Noy, *"there is no single correct ontology for any domain. Ontology design is a creative process and no two ontologies designed by different people would be the same"* (N. Noy et al., 2001). This leads to a natural limitation of our approach as the proposed ontologies are only placeholders for different aspects of instructional design. We have extended and created several ontologies in our ontology framework. However, by definition, every domain can have several perspectives and hence several ontologies. In our ontology framework, we have introduced the notion of meta-ontologies for representing high level aspects of instructional design like process and content, which are then customized for specific cases of Merill's First Principles of Instruction and Bloom's taxonomy. With this context, the following are some potential directions for further research:

- Ontologies for domains beyond adult literacy. The first direction of future work is to apply the proposed framework for school education and skill education. We are currently working on adapting our ontologies to model skill curriculum, specifically focusing on vocational skills.
- Several ontologies other than *goals*, *process*, *content* that were introduced in the ontology framework have to be extended and created in full detail as a natural extension of the framework.
- While we have built tools[12] that help in development of systems, the platform generates *e*Learning Systems specific to adult literacy in India. There is a need for tools that can generate tools to generate tools.
- Creating collaborative, distributed and agile environments for domain and subject matter experts to create, share and disseminate their ontologies is a critical future step towards design of educational technologies for scale and variety.

## Acknowledgements

We would like to thank TCS for providing us with initial inputs for this work, NLM for taking our work forward to create national impact, Government of Telangana for being one of the first adoptors of our technologies and all funding agencies for supporting several international research travels.

[12] http://rice.iiit.ac.in